\documentclass[journal]{IEEEtran}

\usepackage{amsmath, amssymb, bm, cite, psfrag, mathtools}
\usepackage{epstopdf}
\usepackage{rotating}
\usepackage{algorithm,algpseudocode}
\usepackage{array}
\usepackage{supertabular,booktabs}
\usepackage{ragged2e}
\newcolumntype{P}[1]{>{\centering\hspace{0pt}}p{#1}}
\newcolumntype{M}[1]{>{\centering\hspace{0pt}}m{#1}}
\newcolumntype{L}{>{\centering\arraybackslash}m{3cm}}
\usepackage{changepage}
\usepackage[font=small]{caption}
\usepackage{subcaption}
\usepackage{color,soul}
\usepackage{booktabs}

\usepackage{capt-of}
\usepackage{bbm}
\usepackage{multirow}
\usepackage[usenames,dvipsnames]{xcolor}

\usepackage{footnote}
\makesavenoteenv{tabular}
\makesavenoteenv{table}
\usepackage{etoolbox}
\usepackage{pbox}
\usepackage[hyphens]{url}
\usepackage[hidelinks]{hyperref}
\hypersetup{breaklinks=true}
\usepackage{cite}
\usepackage{longtable}
\usepackage{tabu}
\usepackage[margin=0.58in]{geometry}

\usepackage{fixltx2e}
\usepackage{graphicx}
\usepackage{tabu}
\usepackage{booktabs}

\def\dB{\textrm{dB}}

\def\1m{\textrm{1 m}}

\DeclareMathOperator{\sinc}{sinc}

\newtoggle{conference}
\togglefalse{conference} 
\usepackage{tikz}
\usetikzlibrary{calc}
\begin{document}
\bibliographystyle{IEEEtran}

\title{Millimeter-Wave Human Blockage at 73 GHz with a Simple Double Knife-Edge Diffraction Model and Extension for Directional Antennas} 

\author{
George R. MacCartney, Jr.,~\IEEEmembership{Student Member,~IEEE,}
Sijia Deng,~\IEEEmembership{Student Member,~IEEE,}\\
Shu Sun,~\IEEEmembership{Student Member,~IEEE,}
Theodore S. Rappaport,~\IEEEmembership{Fellow,~IEEE}

\thanks{This material is based upon work supported by NOKIA and the NYU WIRELESS Industrial Affiliates Program, three National Science Foundation (NSF) Research Grants: 1320472, 1302336, and 1555332, and the GAANN Fellowship Program. G. R. MacCartney, Jr. (email: gmac@nyu.edu), S. Deng (email: sijia@nyu.edu), S. Sun (email: ss7152@nyu.edu), and T. S. Rappaport (email: tsr@nyu.edu), are with the NYU WIRELESS Research Center, NYU Tandon School of Engineering, Brooklyn, NY 11201.}
}
\maketitle
\begin{tikzpicture}[remember picture, overlay]
\node at ($(current page.north) + (0,-0.25in)$) {G. R. MacCartney, Jr., S. Deng, S. Sun, and T. S. Rappaport, ``Millimeter-Wave Human Blockage at 73 GHz with a Simple Double Knife-Edge};
\node at ($(current page.north) + (0,-0.4in)$) {Diffraction Model and Extension for Directional Antennas," \textit{2016 IEEE 84th Vehicular Technology Conference (VTC2016-Fall)}, Sept. 2016.};
\end{tikzpicture}
\begin{abstract}
This paper presents 73 GHz human blockage measurements for a point-to-point link with a 5 m transmitter-receiver separation distance in an indoor environment, with a human that walked at a speed of approximately 1 m/s at a perpendicular orientation to the line between the transmitter and receiver, at various distances between them. The experiment measures the shadowing effect of a moving human body when using directional antennas at the transmitter and receiver for millimeter-wave radio communications. The measurements were conducted using a 500 Megachips-per-second wideband correlator channel sounder with a 1 GHz first null-to-null RF bandwidth. Results indicate high shadowing attenuation is not just due to the human blocker but also is due to the static directional nature of the antennas used, leading to the need for phased-array antennas to switch beam directions in the presence of obstructions and blockages at millimeter-waves. A simple model for human blockage is provided based on the double knife-edge diffraction (DKED) model where humans are approximated by a rectangular screen with infinite vertical height, similar to the human blockage model given by the METIS project. 
\end{abstract}

\iftoggle{conference}{}{
}

\section{Introduction}\label{sec:intro}
Millimeter-wave (mmWave) bands are an ideal spectrum candidate for fifth-generation (5G) communications systems as the sub-6 GHz spectrum has become overly-saturated~\cite{Andrews14a,Rap13a}. The Federal Communications Commission (FCC) in the United States is actively engaged with industry as the need for mmWave bands becomes more apparent, as indicated by the notice of inquiry (NOI) 14-177~\cite{FCC14177} calling for the investigation of spectrum bands above 24 GHz for mobile use, in addition to the notice of proposed rulemaking (NPRM) 15-138~\cite{FCC15138} for frequency bands in the 27.5-28.35 GHz, 37-38.6 GHz, 38.6-40 GHz, and 64-71 GHz range, for mobile use. Although the FCC only considered the 71-76 GHz and 81-86 GHz E-band frequencies for fixed-services, it is envisioned that they will be considered for mobile use in the future~\cite{Rap16a}. 

The use of multiple antennas to create narrow beams with beamforming technologies will likely be needed for systems operating at mmWaves to maintain acceptable SNR due to the increased free space path loss in the first meter of propagation and reduced diffraction around obstacles, compared to VHF and UHF frequencies~\cite{Andrews14a}. Since mmWave systems are envisaged to employ more directional antennas/antenna systems, they will likely experience greater blockage effects caused by humans, cars, and street furniture, compared to current omnidirectional or quasi-omnidirectional wireless systems~\cite{Rangan14a,Rap15b}. The advent of unlicensed spectrum worldwide in the 60 GHz mmWave~\cite{Pi11a} band has led to technological advancements for commercial grade products due in part to WiGig~\cite{Wigig10a} and IEEE 802.11ad~\cite{Perahia10a}. The 802.11ad 60 GHz channel model~\cite{Maltsev10b} in particular includes a human blockage model to account for random human movements in an environment. 

Human blockage measurements were conducted at 5 GHz by Ericsson to understand the temporal variations induced by humans for stationary terminals in an office over a 200 MHz bandwidth. Results indicated that the line-of-sight (LOS) path was dominant and that secondary paths were 20 dB down or lower~\cite{Medbo04a}. During the day, human blockers on average attenuated the LOS path by 10-15 dB, but no more. A human shadowing model based on double knife-edge diffraction (DKED) was developed with a human scattering model, which were in good agreement with measurement observations. 

Collonge~\emph{et al.} with the Institute of Electronics and Telecommunications of Rennes performed 60 GHz wideband measurements with a 500 MHz bandwidth signal and a 2.3 ns temporal resolution with 40 dB of dynamic range using horn and patch transmitter (TX) and receiver (RX) antennas to capture human activity influence on the channel in a large laboratory~\cite{Collonge04a}. Human shadowing in the direct path between the TX and RX induced more than 20 dB of attenuation for approximately 300 ms for groups of 11 to 15 people walking through the link. Measurements involving horn antennas revealed increased attenuation caused by human blockage compared to the patch antennas due to a lack in angular diversity of directive antennas, an observation noted later in Section~\ref{sec:Results}.

Researchers at IMST GmbH derived a DKED model for human obstruction by use of rays based on 10 GHz measurements from six test cases that modeled human blockers as an infinitely vertical blocking screen with a defined width~\cite{Kunisch08}, represented as a vertical stripe (diffraction screens are commonly used to model buildings in microcellular scenarios as well~\cite{Russell93a}). This method used a local Cartesian coordinate system with a fixed perpendicular orientation between the TX and RX with the screen placed at a defined distance between the two, which created the double-sided knife-edges for modeling purposes. For transmitter-receiver (T-R) separation distances ranging from 5.7 m to 22 m, the maximum single human blockage attenuation was on average 20-30 dB for each distance, with an overall maximum attenuation nearly 40 dB down from the reference. Similar measurements by TU Braunschweig and Intel modeled human blockage through ray-tracing with knife-edge diffraction for the IEEE 802.11ad 60 GHz channel model with good agreement between measurements and the model, and also showed similar deep shadowing attenuation of 30-40 dB while a human blocker walked through a point-to-point link~\cite{Jacob11a}. Additionally, the model proposed in~\cite{Jacob11a} includes a valid model area with an upper and lower bound for blockage attenuation, with an extension to the model given in~\cite{Jacob13a}.

Fraunhofer HHI and TU Braunscheweig also conducted wideband 60 GHz human blockage measurements over a 3 GHz bandwidth with a 2x2 MIMO channel sounder in a conference room for point-to-point links with T-R separation distances of 2, 4, 6, and 10 m with a human blocker walking ``perpendicular walks" (PPWs) through the link~\cite{Peter12a}. DKED, uniform theory of diffraction (UTD), and piecewise linear (PWL) approximation models were compared to the measured data regarding the LOS component and indicated that the DKED model underestimates human body shadowing (HBS) and the UTD model overestimates HBS attenuation. Additional work by Fraunhofer HHI showed that the 60 GHz outdoor channel with human obstructions is highly time-invariant with as much as 30-40 dB attenuation of the LOS path in a typical small cell deployment~\cite{Weiler14a}.

In this paper we present human blockage measurements conducted at 73 GHz for a point-to-point 5 m link using directional high-gain horn antennas in an indoor environment. The rest of the paper is organized as follows: Section~\ref{sec:Meas} describes the measurement equipment and setup, Section~\ref{sec:Model} describes a modified METIS model based on DKED for simulation and measurement comparison, Section~\ref{sec:Results} compares the modified DKED model simulations and the measured results, with conclusions drawn in Section~\ref{sec:Conclusion}. 

\section{Measurement Equipment and Descriptions}\label{sec:Meas}
\subsection{Measurement Equipment and Specifications}
The 73 GHz human blockage measurements were conducted with a wideband correlator channel sounder that transmits a 500 Megachips-per-second pseudorandom noise (PN) sequence of length 2047, resulting in a 1 GHz null-to-null RF bandwidth centered at 73.5 GHz with a transmit power of -5.8 dBm. The TX and RX employed a high-gain directional horn antenna each with a 15$^\circ$ azimuth and elevation half-power beamwidth (HPBW) and 20 dBi of gain, resulting in an effective isotropic radiated power (EIRP) of 14.2 dBm. The system employed a National Instruments (NI) 5771 high-speed digitizer that sampled at 1.5 Giga-Samples/s (GS/s) on the $I$ and $Q$ demodulated baseband channels that were captured and correlated in software via a fast Fourier transform (FFT) matched filter to create the $I$ and $Q$ channel impulse responses (CIRs) that generate a power delay profile (PDP) of the channel ($I^2+Q^2$). The system had a maximum instantaneous dynamic range of 40 dB, with a 25 dB down from max peak PDP threshold applied to each individual PDP to discard erroneous noise spikes and system irregularities near the noise floor. The minimum periodic PDP interval for capture is 32.752 $\mu$s, with up to 41,000 periodic PDPs recorded per snapshot, where a snapshot is the recording of successive PDPs. Note that the TX and RX channel sounding systems shared a common cable-connected 10 MHz reference for frequency synchronization. Additional details regarding the measurement system are provided in Table~\ref{tbl:MeasSys}.
\begin{table}
	\centering
	\caption{Wideband correlator channel sounder system specifications for the 73 GHz human blockage measurements.}
	\label{tbl:MeasSys}
	\begin{center}
		\scalebox{0.82}{
		\begin{tabu}{|c|[1.6pt]c|}
			\hline 
			Description &	Specification \\ \specialrule{1.5pt}{0pt}{0pt}
			Baseband Sequence &	PRBS (11th order: 2$^{11}$-1 =  Length 2047)	\\ \hline
			Chip Rate &	500 Mcps	\\ \hline
			RF Null-to-Nulll Bandwidth &	1 GHz	\\ \hline
			PDP Detection &	FFT matched filter 	\\ \hline
			Sampling Rate &	1.5 GS/s on $I$ and $Q$	\\ \hline
			Multipath Time Resolution &	2 ns	\\ \hline
			Minimum Periodic PDP Interval &	32.752 $\mu$s	\\ \hline
			Maximum Periodic PDP records per snapshot &	41,000 PDPs	\\ \hline
			PDP Threshold &	25 dB down from max peak	\\ \hline
			TX/RX Intermediate Frequency &	5.625 GHz	\\ \hline
			TX/RX LO &	67.875 GHz (22.625 GHz x3)	\\ \hline
			Synchronization &	TX/RX Share 10 MHz Reference	\\ \hline
			Carrier Frequency &	73.5 GHz	\\ \hline
			TX Power &	-5.8 dBm	\\ \hline
			TX/RX Antenna Gain &	20 dBi	\\ \hline
			TX/RX Azimuth and Elevation HPBW &	15$^\circ$	\\ \hline
			TX/RX Antenna Polarization &	V-V	\\ \hline
			EIRP &	14.2 dBm	\\ \hline
			TX/RX Heights &	1.4 m	\\ \hline
		\end{tabu}}
	\end{center}
\end{table}
\subsection{Measurement Environment and Descriptions}
\begin{figure}
	\centering
	\begin{subfigure}[b]{0.5\textwidth}
		\includegraphics[width=1\linewidth]{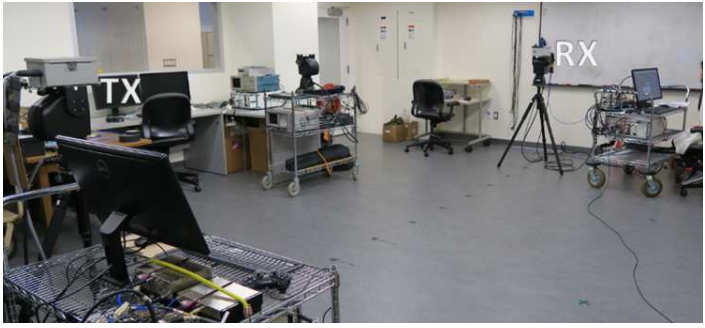}
		\caption{Side view of TX and RX.}
		\label{fig:TXRX_open} 
	\end{subfigure}
	\begin{subfigure}[b]{0.5\textwidth}
		\includegraphics[width=1\linewidth]{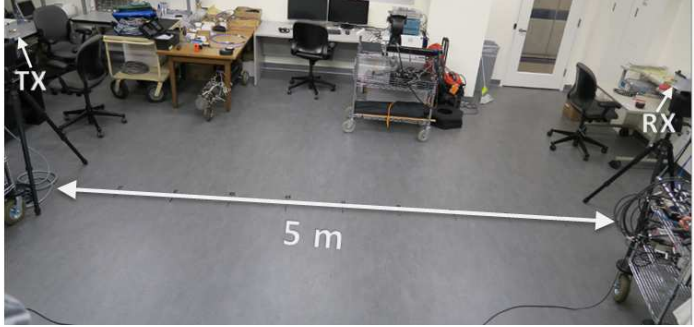}
		\caption{Top-down view of TX and RX with 5 m separation distance.}
		\label{fig:TXRX_5m}
	\end{subfigure}
	\caption{(a) Side view of TX and RX. (b) Top-down view of the TX and RX.}
\end{figure}

\begin{figure}
	\centering
	\includegraphics[width=0.5\textwidth]{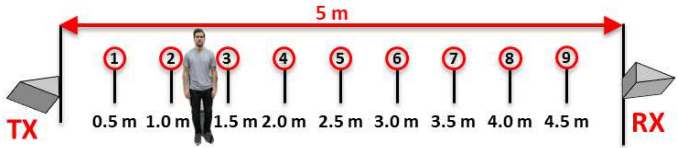}
	\caption{Depiction of nine measurements where the human blocker walked through the LOS link at a perpendicular orientation (side of body) at 0.5 m increments between the TX and RX.}
	\label{fig:Test4Env}
\end{figure}

\begin{figure}
	\centering
	\includegraphics[width=0.5\textwidth]{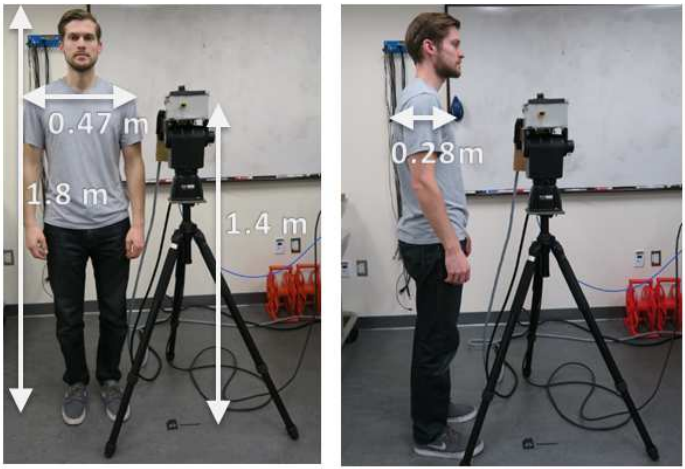}
	\caption{Human blocker dimensions with the TX and RX. $b_{\text{height}}$ = 1.8 m; $b_{\text{breadth}}$ = 0.47 m; $b_{\text{depth}}$ = 0.28 m; TX/RX height = 1.4 m.}  
	\label{fig:BodyPic}
\end{figure}

The TX and RX were set up in an open laboratory space with a 5 m T-R separation distance with each antenna height at 1.4 m and oriented boresight-to-boresight. Figs.~\ref{fig:TXRX_open} and~\ref{fig:TXRX_5m} show the side and top-down environment views. 
Nine separate, 5-second measurement snapshots were recorded for a human blocker walking through the point-to-point link, with a PDP interval frequency of 500 Hz, resulting in approximately 2500 PDPs per snapshot (1 PDP captured every 2 ms, with no averaging). Measurements 1 through 9 were conducted with the human blocker walking (approximately 1 m/s) at a perpendicular orientation through the LOS link at discrete 0.5 m increments between the TX and RX from 0.5 m to 4.5 m\footnote{The Fraunhofer distance of the antennas is 0.2 m, thus 0.5 m is in the far-field.} as depicted in Fig.~\ref{fig:Test4Env}. The body breadth of the blocker (shoulder-to-shoulder) is $b_{\text{breadth}}$ = 0.47 m, the depth is $b_{\text{depth}}$ = 0.28 m (similar to the dimensions in~\cite{Peter12a}), and the height is $b_{\text{height}}$ = 1.8 m with a sketch provided in Fig.~\ref{fig:BodyPic} to show comparison of the blocker dimensions with the TX and RX. For the measurement results given in Section~\ref{sec:Results}, only a portion of the 5-second snapshot for each measurement is displayed in order to convey the relevant shadowing events (SE). 
\section{Double Knife-Edge Diffraction Blockage Model}\label{sec:Model}
\begin{figure}
	\centering
	\begin{subfigure}[b]{0.5\textwidth}
		\includegraphics[width=1\linewidth]{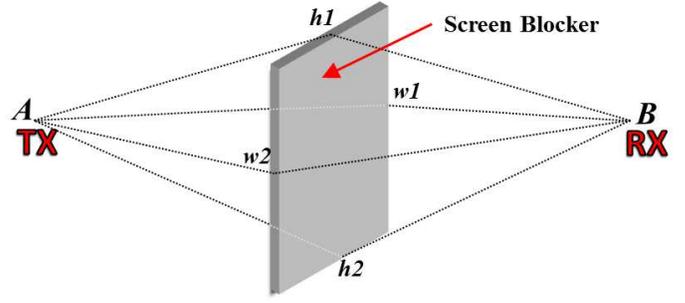}
		\caption{3D screen projection.}
		\label{fig:3DProj} 
	\end{subfigure}
	\begin{subfigure}[b]{0.5\textwidth}
		\includegraphics[width=1\linewidth]{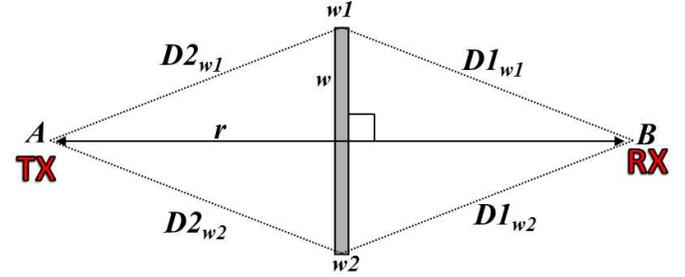}
		\caption{Top-down screen projection.}
		\label{fig:TopProj}
	\end{subfigure}
	\begin{subfigure}[b]{0.5\textwidth}
		\includegraphics[width=1\linewidth]{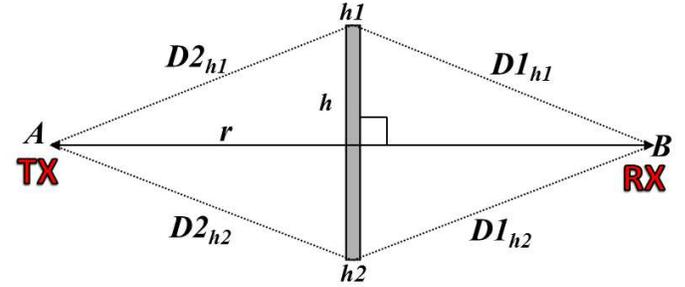}
		\caption{Side screen projection.}
		\label{fig:SideProj}
	\end{subfigure}
	\caption{(a) 3D projection of screen blocker. (b) Top-down projection of screen blocker. $r$ is the distance between the TX (point $A$) and RX (point $B$), and $w$ is the width of the screen. (c) Side projection of screen blocker. $r$ is the same distance between the TX and RX as in the top-down view, and $h$ is the height of the screen. The dimensions are defined as follows:
		$r\overset{\text{def}}{=}\overline{AB}$;
		$w\overset{\text{def}}{=}\overline{w1w2}$;
		$h\overset{\text{def}}{=}\overline{h1h2}$;
		$D2_{w1}\overset{\text{def}}{=}\overline{A w1}$; $D1_{w1}\overset{\text{def}}{=}\overline{w1 B}$; $D2_{w2}\overset{\text{def}}{=}\overline{A w2}$; $D1_{w2}\overset{\text{def}}{=}\overline{w2 B}$;
		$D2_{h1}\overset{\text{def}}{=}\overline{A h1}$; $D1_{h1}\overset{\text{def}}{=}\overline{h1 B}$; $D2_{h2}\overset{\text{def}}{=}\overline{A h2}$; $D1_{h2}\overset{\text{def}}{=}\overline{h2 B}$.}
\end{figure}

A simple DKED model assumes a human blocker to be represented as a screen with four sides, or as an infinitely vertical screen with two sides~\cite{Kunisch08,Medbo13a}. A numerical approximation for DKED from~\cite{Medbo13a} was developed in the METIS human blockage model~\cite{METIS2015} and is the model used for comparison and extension in this paper. A sketch of the screen used to represent a human for the DKED model is shown in Figs.~\ref{fig:3DProj},~\ref{fig:TopProj}, and~\ref{fig:SideProj} for the 3D, top-down, and side projections, respectively, with dimensions (all in meters) of each length defined in the caption. 

Considering Figs.~\ref{fig:3DProj},~\ref{fig:TopProj}, and~\ref{fig:SideProj} given here, shadowing due to the human blocker is determined for a screen blocker perpendicular to the TX-RX pointing orientation, with dimensions of the blocker that represent the general depth ($b_{\text{depth}}$) and height ($b_{\text{height}}$) of the blocker, with respect to the solid lines with arrows that connect the TX (point $A$) and RX (point $B$) in the top-down and side view in Figs.~\ref{fig:TopProj} and~\ref{fig:SideProj}, respectively. As the screen moves between the TX and RX, it remains perpendicular to the point-to-point link for simple simulation and analysis as explained in~\cite{METIS2015}. The corresponding shadowing incurred by each of the 4 edges ($w1$; $w2$; $h1$; $h2$) based on knife-edge diffraction is as follows~\cite{METIS2015}:
\begin{subequations}
	\begin{equation}\label{eq:Fw1}
	F_{w1|w2} = \frac{\tan^{-1}\left(\pm\frac{\pi}{2}\sqrt{\frac{\pi}{\lambda}(D2_{w1|w2}+D1_{w1|w2}-r}\right)}{\pi}
	\end{equation}
	\begin{equation}\label{eq:Fh1}
	F_{h1|h2} = \frac{\tan^{-1}\left(\pm\frac{\pi}{2}\sqrt{\frac{\pi}{\lambda}(D2_{h1|h2}+D1_{h1|h2}-r}\right)}{\pi}
	\end{equation}
\end{subequations}
where $F_{w1|w2} = F_{w1} \text{ or } F_{w2}$, $\lambda$ is the carrier wavelength, $D2$ and $D1$ are the projected distances according to the dimensions from the top-down and side views between the TX and the screen and between the screen and the RX, respectively, with $r$ as the T-R separation distance in the side view and top-down view. The width of the screen is defined as $w$ from Fig.~\ref{fig:TopProj} and the height is defined as $h$ from Fig.~\ref{fig:SideProj}. The $\pm$ sign is applied with a $+$ to both edges for non-LOS (NLOS) shadow zone conditions, and for LOS conditions, the edge farthest from the link is considered in the shadow zone and is applied with a $+$, and the edge closest to the link is applied with a $-$. For a situation with multiple screens, the total loss is determined by summing the combined losses in dB. The advantage to this model is the simplified and closed-form numerical approximation for DKED, similar to solutions used to model UTD for real-time propagation prediction modeling~\cite{Wang05b}. 

When considering a blocker with a given height $h$ and width $w$, the shadowing loss by a blocker is modeled by DKED for a screen with four edges ($w1$; $w2$; $h1$; $h2$) as~\cite{METIS2015}:
\begin{equation}\label{eq:screenWH}
	L_{\text{screen}}[\dB] = -20\log_{10}\left(1-(F_{h1}+F_{h2})(F_{w1}+F_{w2})\right)
\end{equation}
with the corresponding four losses due to knife-edge diffraction from the sides, top, and bottom of the screen. However, a simpler model considers only the side edges for diffraction, and considers the screen height $h$ to be infinitely vertical:
\begin{equation}\label{eq:screenW}
L_{\text{screen}}[\dB] = -20\log_{10}\left(1-(F_{w1}+F_{w2})\right)
\end{equation}
The models in~\eqref{eq:screenWH} and~\eqref{eq:screenW} are accurate for omnidirectional antenna measurements, but do not account for the lack of angular diversity inherent in measurements with directive antennas. Thus, the antenna radiation pattern as pointed out in~\cite{Collonge04a} has an impact on the blockage attenuation. In order to account for the antenna radiation pattern at off-boresight angles that influences diffraction gain, the KED fields must be weighted accordingly. That is, when the screen is fully blocking the LOS path (boresight-to-boresight is blocked), the projected paths between the TX and screen ($D2_{w1}$ and $D2_{w2}$), and the screen and the RX ($D1_{w1}$ and $D1_{w2}$), do not contain the full directive gain of the antennas. 

\begin{figure*}[ht!]
	\centering
	\includegraphics[width=6.7in]{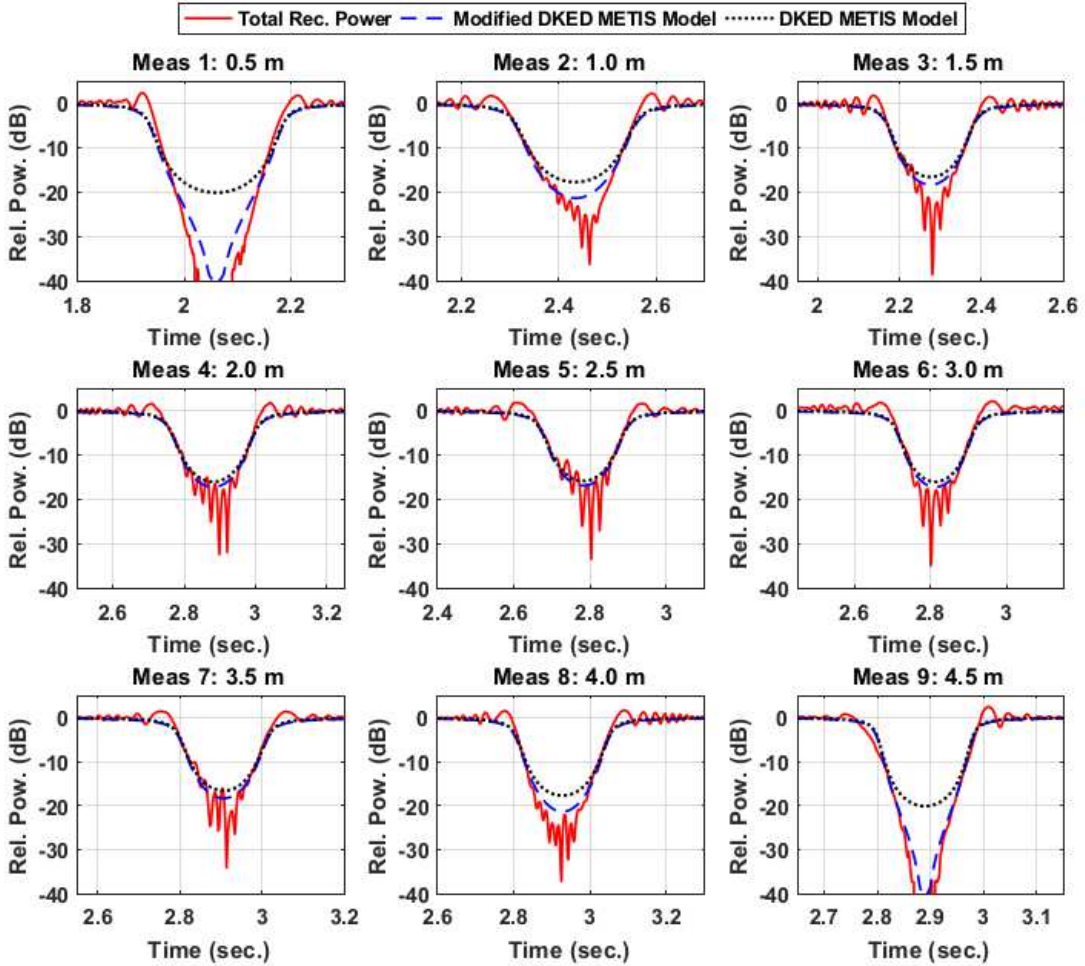}
	\caption{Comparison of total received power from human blockage measurements at 73 GHz and DKED simulations for an infinitely vertical screen with width 0.28 m.}
	\label{fig:DKE_Sim}
\end{figure*}

The normalized azimuth gain $G$ at angle $\theta$ is determined via the far-field radiation pattern of a horn antenna with a specific azimuth half-power beamwidth ($\text{HPBW}_{\text{AZ}})$ and is approximated by~\cite{Sun15a}:
\vspace{0.2cm}
\begin{align}
G(\theta) = \sinc^2(\textrm{a}\cdot \sin (\theta))\cdot \cos^2(\theta)
\end{align}
where:
\begin{align}
\sinc^2\Bigg(\textrm{a}\cdot \sin \bigg( \frac{\text{HPBW}_{\text{AZ}}}{2}\bigg)\Bigg)\cdot \cos^2 \bigg( \frac{\text{HPBW}_{\text{AZ}}}{2}\bigg)=\frac{1}{2}
\end{align}
A modified version of~\eqref{eq:screenW} that accounts for the TX and RX antenna radiation pattern is given by:
\begin{equation}\label{eq:modMet}
\begin{split}
L_{\textrm{Screen Mod.}}[\dB] = -20\log_{10}\Bigg \lvert \bigg(\frac{1}{2}-F_{w1}\bigg)\cdot \sqrt{G_{D2_{w1}}}\cdot \sqrt{G_{D1_{w1}}} \\+\bigg(\frac{1}{2}-F_{w2}\bigg) \cdot \sqrt{G_{D2_{w2}}}\cdot \sqrt{G_{D1_{w2}}}\Bigg \rvert
\end{split}
\end{equation}
where $G_{D2_{w1}|D1_{w1}|D2_{w2}|D1_{w2}}$ are the normalized linear gains of the directional TX and RX antennas based on the projected angle from the TX to the screen ($D2_{w1|w2}$) and from the screen to the RX ($D1_{w1|w2}$), relative to the normalized boresight gain which has $G(0)=1$. This results in larger losses in the shadow region for what would be observed for an omnidirectional antenna, due to off boresight trajectories. Note that when the screen does not shadow the boresight trajectory between the TX and RX, the normalized gains are set to $G =1$.

\section{Measurement Results and Analysis}\label{sec:Results}
The measurement setup described in Section~\ref{sec:Meas} indicated that 500 PDPs were captured each second for a total of 5 seconds (approximately 2500 PDPs) for each of the nine measurement distances between the TX and RX (see Fig.~\ref{fig:Test4Env}) as the human blocker walked along a path that is perpendicular to the line between the TX and RX. The area under each PDP was summed up to determine the total received power (2500 total samples in 5 seconds) due to contributions from the LOS path, and reflected and diffracted paths (a 25 dB down from max peak threshold was used for each individual PDP). In order to consistently account for the human blocker moving through the LOS link, an unobstructed PDP was used to obtain the reference received power in the absence of a shadowing event. Thus, all powers were referenced to an unobstructed power of 0 dBm, with subsequent human shadowing gain observed as positive gain in dB and human shadowing attenuation observed as negative gain in dB.

For each of the nine measurement cases, the DKED METIS screen model~\eqref{eq:screenW} and modified DKED METIS model~\eqref{eq:modMet} were simulated based on the approximate 1 m/s walking speed of the human blocker and a depth ($b_{\text{depth}}$) of 0.28 m as the width $w$ for the vertical screen. Fig.~\ref{fig:DKE_Sim} shows the measured total received power and simulated DKED models for each of the nine measurements. Observations from each measurement indicate that a diffraction/reflection gain caused by the human blocker upon entering the shadow region can induce a gain of 1-2 dB in a majority of the cases, irrespective of the distance at which the blocker entered the link. 

The maximum blockage attenuation was greater than 30 dB in the shadow region for all nine measurement cases. The measured results indicate a strong sensitivity to human blockage when in the shadow region between the TX and RX, and the small wavelength of the signal (approximately 4 mm) attributes to the deep oscillating fades of 30 to 40 dB or greater, as a result of variations in constructive and destructive interference. Similarly, the top envelope of the human blockage attenuation can be attributed to constructive interference. When the blocker is very close to the TX or RX (Meas. 1-2, and Meas. 8-9), the DKED METIS screen model significantly underestimates the observed shadowing, in extreme cases by more than 20 dB (Meas. 1 and Meas. 9) and by approximately 10 dB for Meas. 2 and Meas. 8. 

The use of directive antennas had a significant impact on human blockage measurement results, especially when the blocker was close to the TX and RX where attenuation was 40 dB or greater in the deep shadow region. This is a strong implication on the use of directive antennas and beamforming at mmWaves where such narrow beams with high gain and fast switching speeds will be needed to improve SNR and to avoid blockers~\cite{Sun14b}. While the original DKED METIS model was unable to effectively model a human blocker (modeled as an infinitively high screen), the modified DKED METIS model accounted for the 15$^\circ$ HPBW antennas used at the TX and RX (see Table~\ref{tbl:MeasSys}), and was able to model the screen with good agreement between measurements and simulation as depicted in Fig.~\ref{fig:DKE_Sim}. By accounting for antenna directivity at the TX and RX, the model accurately traces the upper envelope of the blockage attenuation in all measurement cases. In each of the nine measurement cases, the entire shadowing event lasted approximately 200 ms to 300 ms on average (nominal 0 dB attenuation to maximum attenuation, and back to nominal 0 dB attenuation), with 20 dB to 40 dB shadowing events in most cases. The average walking speed of 1 m/s for the human blocker and the body depth ($b_{\text{depth}}$) of 0.28 m corresponds to a shadowing event of approximately 280 ms, well aligned with the observations. This temporal variation in power could be overcome by beams switching directions to find scatterers and reflections to make a reliable connection~\cite{Andrews14a,Rap13a,Sun14b,Azar13a}. Shadowing event durations and fade depths were symmetrical when using equivalent TX and RX directional antennas, such that nearly identical observations were made when the human body was blocking the LOS path at a distance of 0.5 meters from the TX (Meas. 1) and 0.5 meters from the RX (Meas 9.), and similarly for 1 meter from the TX (Meas. 2) and 1 meter from the RX (Meas. 8), as shown in Fig.~\ref{fig:DKE_Sim}.

The DKED simulations in this paper did not consider the height edges of the screen (see Fig.~\ref{fig:3DProj}), which has also been commonly ignored by others in the literature~\cite{Kunisch08,METIS2015,Medbo04a}. The UTD model was not considered but was observed to overestimate human blockage in~\cite{Peter12a} in addition to the DKED model underestimating human blockage. It is expected that antennas with a wider beamwidth would result in less blockage attenuation that would more closely match the original DKED METIS model, and thus an antenna beamwidth dependent extension was added (e.g. Eq.~\eqref{eq:modMet} herein) to the modified DKED blockage model with simulation and measurement validation. This simple diffraction model can be computed in software and added to a ray-tracer to simulate human blockage and similar obstructions and their losses as part of a partition loss model~\cite{Skidmore96a}. Furthermore, the DKED models in this paper and in METIS~\cite{METIS2015} can be extended to a more complicated model with phase correction and non-perpendicular screen orientations~\cite{Kunisch08}, although this is unnecessary, given the good agreement between the model simulations and measurements presented in this paper and others in the literature. 

\section{Conclusion}\label{sec:Conclusion}
This paper presented some of the first published work on 73 GHz E-band human blockage measurements. Measurement results for various perpendicular paths between a TX and RX separated by 5 m indicated that an average sized male with a body depth of 0.28 m can cause as much as 30-40 dB of attenuation or more on average, with the use of directional antennas. A simple DKED model and a modified DKED model for an infinitely vertical screen were used to compare model simulations and measurements, with good agreement observed between the measurements and the modified DKED METIS model that accounted for antenna directivity. When using identical TX and RX directional antennas, shadowing event durations and fade depths were observed to be symmetrical when the human blocker walked through the LOS link at equivalent distances from the TX and RX (e.g. Meas. 1 and Meas. 9 from Fig.~\ref{fig:DKE_Sim}). Large shadowing events (30 dB to 40 dB of attenuation) that lasted approximately 200 ms to 300 ms for a single blocker may be overcome by using beamforming and phased-array architectures to switch beam directions to find reflections and scatterers in order to avoid blockages and obstructions, in the 5G era. Future work will investigate different human blockage scenarios and the constructive and destructive effects observed for shadowing events caused by human blockers.
\bibliography{VTC_f2016_Blockage_v3_0}
\bibliographystyle{IEEEtran}

\end{document}